# Topological Updating Schemes: A Case Study In 3-d U(1)

Arjan Hulsebos[a] and Peer Ueberholz[b]

[a]FB08, University of Wuppertal, Gaußstraße 20, 42119 Wuppertal, BRD.
[b]Dept. of Physics and Astronomy, University of Edinburgh, King's Building, Edinburgh, EH9 3JZ, Scotland.

We study a topological updating scheme in three dimensional U(1) gauge theory. Some expectations for four dimensional SU(N) gauge theories are discussed.

## 1. Introduction

The presence of topological objects in a lattice theory may play havoc with updating schemes. The problem is that changes in the number of these objects usually involve fairly large changes in quite a number of lattice variables, resulting in large changes in the action, and hence in a very low acceptance rate. In other words, there is a very poor sampling over the different topological sectors, characterized by the number of said objects. Not only in theories that contain objects of a genuine topological nature, such as instantons and topological charge, do these problems arise, but also in theories that possess quasi-topological objects, such as monopoles and vortices.

A strategy to tackle this problem is to devise an algorithm that proposes these changes with a large probability of acceptance. The cluster algorithm succeeds in doing just that for, e.g., the XY models[1]. When this proves to be too difficult, e.g. for SU(N) gauge theories in four dimensions, we may consider 'adding' transitions by hand[2]. The question here is how to get a reasonable acceptance probability for these 'topological updates'. By reasonable we mean larger than, say, 5%. We will address this problem in 3-d U(1) pure gauge theory.

## 2. The model and the method

The Wilson action for 3-d U(1) gauge theory is given by

$$S = \beta \sum_{x,\mu,\nu} \text{Re}\, U_{\mu\nu}(x) \qquad (1)$$

In this theory, we can define monopoles in the following way. Writing $U_{\mu\nu}(x) = \exp(i\,\theta_{\mu\nu}(\underline{x}))$, we can split $\theta$: $\theta_{\mu\nu}(x) = \bar{\theta}_{\mu\nu}(x) + 2\pi n_{\mu\nu}(x), \bar{\theta} \in \langle -\pi, \pi]$, n integer. Note that n = $0, \pm 1, \pm 2$. We can transpose this to the dual lattice by $j_\mu(\hat{x} - \hat{\mu}) = \frac{1}{2}\varepsilon_{\mu\nu\rho} n_{\nu\rho}(x)$, which satisfies the following relation (dropping the hat from now on):

$$\begin{aligned} \partial'_\mu j_\mu(x) &= m(x) \\ m(x) &= 0, \pm 1, \pm 2, \pm 3. \end{aligned} \qquad (2)$$

When $m(x) > 0 (< 0)$, there is a (anti-) monopole at site $x$ on the dual lattice. When $m(x) = 0$, but any of the six $j$'s connected to $x$ is non-zero, a Dirac string passes through $x$. It is easy to see that then an even number of $j$'s are non-zero. Nothing prevents $j$ from forming closed loops. This is indeed what is found[3]. We can now distinguish between two types of loops: ones that are closed through the boundary, and ones that are not. Although we can assign a winding number per direction to former type, it turns out that, in each direction, only the total sum of winding numbers is gauge invariant. It is possible to split a single loop with winding number 2 into two loops with winding number 1, merely by performing a gauge transformation.

These Dirac strings introduce meta-stability[4, 5]. Suppose we have a gauge configuration with a single Dirac string with winding number 1. In order to change this number, we need to create another string. As this involves changes over large distances, the probability of this happening will be (very) low in, e.g., a standard Metropolis algorithm.

If we now take a slice of the lattice through which a Dirac string pierces, and look upon this as a *two dimensional* gauge configuration, we find that the topological charge of this 2-d configuration is equal to the winding number of this string, i.e. ±1. So we now have a scheme to make a topological update for 3-d U(1): we simply add a stack of 2-d U(1) configurations with topological charge +1 (or -1) to our 3-d U(1) configuration, and perform a global Metropolis on this new configuration.

There are different types of 2-d configurations we can use. We can use configurations from an equilibrium ensemble at a given $\beta_{2\text{-d}}$. We can also construct gauge configurations with constant field strength. We will use 2-d gauge configurations from several $\beta_{2\text{-d}}$ values, as well as specially constructed configurations:

$$U_1(x) = \exp(i\omega(x_2 - 1)),$$
$$U_2(x) = \mathbf{1} \quad \text{for } x_2 = 1,\ldots,(N-1)$$
$$U_1(x) = \exp(i\omega(x_2 - 1)),$$
$$U_2(x) = \exp(-i\omega N x_1)$$
$$\text{for } x_2 = N \quad (3)$$

where $\omega = 2\pi/N^2$. This leads to a gauge configuration with a field strength of $U_P(x) = \exp(-i\omega)$ throughout the lattice. The topological charge for this configuration is $Q = -1$.

From a theoretical point of view we can make an estimate of the acceptance rate of these topological updates. From [6], we find

$$\Omega = \text{erfc}\,(\frac{1}{2}\sqrt{h_1}), \qquad h_1 \equiv <\Delta\mathcal{S}>,$$
$$\text{where } \Delta\mathcal{S} \equiv S(U_{\mu\nu}) - S(U_{\mu\nu} + U_{2\text{-d}}). \quad (4)$$

From this, we expect the following,

| type | $\Delta\mathcal{S}$ | $\Omega$ |
|---|---|---|
| eq. (3) | $\beta <U_P> \frac{2\pi^2}{N^2}$ | 0.252 |
| 2-d conf. | $\beta N^3 <U_P>[1- <U_P^{2d}>]$ | |

for a $16^3$ lattice at $\beta$ = 2.50, for which $<U_P>$=0.8542(6). $<U_P^{2d}>$ >0.9 for $\beta^{2d} > 4.0$.

## 3. Results

So from this analysis we expect no topological updates for any 2-d gauge configurations but the one with a constant field strength. This is indeed what is found. We made 15k updates, being 5-hit Metropolis, after discarding 2k updates. Dirac strings were proposed every 10 updates. We indeed found that no topological updates were accepted for any 2-d gauge configurations but the one with a constant field strength. Here, we found an acceptance rate of 0.133(17), which is remarkably smaller than the 'expected' 0.252.

| $\beta$ | acc. rate | $<U_P>$ | $\Omega$ | $M$ |
|---|---|---|---|---|
| 1.50 | 0.436(42) | 0.6880(5) | 0.4249 | 222(2) |
| 1.75 | 0.374(21) | 0.7619(4) | 0.3644 | 75.0(1.5) |
| 2.00 | 0.334(37) | 0.8061(3) | 0.3186 | 22.7(1.0) |
| 2.25 | 0.316(21) | 0.8342(5) | 0.2819 | 6.6(1.0) |
| 2.375 | 0.164(44) | 0.8453(2) | 0.2658 | 1.9(2) |
| 2.50 | 0.152(45) | 0.8542(1) | 0.2516 | 0.97(15) |
| 2.75 | 0.128(45) | 0.8700(1) | 0.2244 | 0.21(2) |
| 3.00 | 0.096(33) | 0.8809(1) | 0.2017 | 0.054(8) |
| 3.25 | 0.032(12) | 0.8909(1) | 0.1814 | 0.015(4) |
| 3.50 | 0.056(31) | 0.8992(1) | 0.1635 | 0.004(2) |
| 3.75 | 0.040(25) | 0.9036(1) | 0.1482 | — |
| 4.00 | 0.024(16) | 0.9125(1) | 0.1335 | — |
| 4.25 | 0.024(10) | 0.9180(1) | 0.1208 | — |
| 4.50 | 0.020(8) | 0.9228(1) | 0.1095 | — |
| 4.75 | 0.012(16) | 0.9270(1) | 0.0994 | — |
| 5.00 | 0.008(11) | 0.9309(1) | 0.0902 | — |

Table 1 : The results for 5k measurements per $\beta$ value.

Returning to 2-d gauge configurations from an equilibrium ensemble; we might consider choosing a very large $\beta^{2\text{-d}}$. This will lead to an average plaquette, which will be very close to unity, hence rendering a reasonable acceptance rate. However, since such a two dimensional gauge field is required to carry a topological charge $Q = \pm 1$, we will find that all plaquettes but one or two will be very close to unity, while the remaining one or two will be close to *minus* unity. For these configurations we get $\Delta\mathcal{S} = 4\beta <U_P> = 8.542$ and thus $\Omega < 10^{-8}$.

Being surprised by the large deviation of the acceptance rate we found at $\beta = 2.50$ from the theoretical one, we made shorter runs at several $\beta$ values to check the validity of our formula (4). The results are as follows:

The transition around $\beta = 2.3$ is striking. It would appear that Dirac strings become suppressed for larger $\beta$ values. This becomes very reasonable when we consider the average monopole number, $M = < \sum_x |m(x)| >$, presented in the last column of table 1.

So the transition in the acceptance rate occurs where the average number of monopoles in a configuration drops below 2. It is then not at all surprising that Dirac strings also become (very) rare.

## 4. Outlook for SU(N)

The requirement for topological updates to have a reasonable acceptance rate turns out to be a very restrictive one on the actual implementation of the algorithm. As a matter of fact, only the smoothest updates possible stand a chance of having a reasonable acceptance rate. We also found that there can be large deviations from the theoretical predictions, due to the intrinsic properties of the model, which have not been accounted for in the analysis.

Let us now speculate on 4-d SU(N) gauge theories. As all these theories allow for topological charge and instantons, we expect very large tunnelling times from one topological charge sector to another for sufficiently large gauge coupling $\beta$. As we have seen, taking a realistic gauge configuration will lead us nowhere, so we have to resort to specially constructed gauge configurations.

Taking SU(2) for the moment, we can construct two types of configurations. The first one has constant field strength for each $(\mu, \nu)$ plane, and is given by

$$U_\mu(x) = \exp(i\tau \sum_{\nu \neq \mu} [\omega_{\mu\nu} x_\nu (1 - \delta_{x_\nu,(L_\nu-1)}) - L_\nu \omega_{\mu\nu} x_\mu \delta_{x_\nu,(L_\nu-1)}]) \quad (5)$$

where, from the co-cycle condition $\omega_{\mu\nu} = 2\pi n_{\mu\nu}/L^2$, $n_{\mu\nu}$ integer. Note that $n_{\mu\nu} \neq n_{\nu\mu}$ is allowed. $\tau$ is one of the Pauli matrices, or possibly a properly normalized linear combination of them. This configuration carries topological charge $Q = 2[(n_{12} - n_{21})(n_{34} - n_{43}) - (n_{13} - n_{31})(n_{24} - n_{42}) + (n_{14} - n_{41})(n_{23} - n_{32})]$ Notice that $Q$ is even. The other configuration is the well-known single instanton configuration[7–9].

These configurations all have rather large actions. The one with constant field strengths has an action of $S/\beta = 2\pi^2 \sum_{\mu\nu} (n_{\mu\nu} - n_{\nu\mu})^2$, while the single instanton configuration has an action of $S/\beta = 9.6$, but which can be lowered to $S/\beta = 6.8$ by cooling[8].

These numbers are not encouraging from the topological update's point of view, but it should be kept in mind that for non-Abelian theories, $<\Delta S>$ may turn out to be smaller than expected from this analysis.

## Acknowledgements

This study started from a discussion by one of us with Martin Grabenstein. We aould like to thank him also for useful e-mail exchanges and further discussions.